\documentclass[aps,nofootinbib,preprintnumbers,twocolumn,superscriptaddress,prd]{revtex4-1}

\usepackage{amsmath}
\usepackage{amssymb}
\usepackage{slashed}
\usepackage{epsfig}
\usepackage{bbm,bm}
\usepackage{hyperref}
\usepackage{color}
\usepackage{comment}
\usepackage{amsfonts}
\usepackage{dsfont}
\usepackage{multirow}
\usepackage{tensor}
\usepackage[normalem]{ulem}
\usepackage{xcolor,colortbl}

\def\bea{\begin{eqnarray}} 
\def\eea{\end{eqnarray}}
\def\be{\begin{equation}} 
\def\ee{\end{equation}} 
\def\ba{\begin{array}}
\def\ea{\end{array}}

\def\be{\begin{equation}}
\def\ee{\end{equation}}
\def\bea{\begin{eqnarray}}
\def\eea{\end{eqnarray}}

\definecolor{Gray}{gray}{0.9}
\newcolumntype{a}{>{\columncolor{Gray}}c}

\let\oldtitle\title
\renewcommand{\title}[1]{\oldtitle{\color{blue}{#1}}}

\begin{document}

\title{
Crossover exponents, fractal dimensions and logarithms\\
in Landau-Potts field theories
}

\author{M.\ Safari}
\email{mahsafa@gmail.com}
\affiliation{Romanian Institute of Science and Technology, 
Str.~Virgil Fulicea  3, 400022 Cluj-Napoca, Rom\^ania}

\author{G.\ P.\ Vacca}
\email{vacca@bo.infn.it}
\affiliation{INFN - Sezione di Bologna, via Irnerio 46, I-40126 Bologna, Italy}

\author{O.\ Zanusso}
\email{omar.zanusso@unipi.it}
\affiliation{Universit\`a di Pisa and INFN - Sezione di Pisa, Largo Bruno Pontecorvo 3, I-56127 Pisa, Italy}

\begin{abstract}
%
We compute the crossover exponents of all quadratic and cubic deformations
of critical field theories with permutation symmetry $S_q$ in
$d=6-\epsilon$ (Landau-Potts field theories)
and $d=4-\epsilon$ (hypertetrahedral models)
up to three loops.
We use our results to determine the $\epsilon$-expansion of
the fractal dimension of critical clusters in the most interesting cases,
which include spanning trees and forests ($q\to0$), and bond percolations ($q\to1$).
We also explicitly verify several expected degeneracies in the spectrum
of relevant operators for natural values of $q$ upon analytic continuation,
which are linked to logarithmic corrections of CFT correlators,
and use the $\epsilon$-expansion to determine the universal coefficients
of such logarithms.
\end{abstract}

\pacs{}

\maketitle

\section{Introduction}

The lattice Potts model can be seen as a generalization of the simple ferromagnetic Ising model at zero magnetic field,
in which the two possible spin states, up and down, are replaced by $q$ distinct states which interact locally
through a term that favors the same state between neighboring sites.
By construction, the Potts model is invariant under the permutation group $S_q$ of the states, which recovers
the symmetry of the Ising model for $q=2$ when, in fact, $S_2 \simeq \mathbb{Z}_2$ \cite{1976JPhA....9.1441A}.
The similarity with the Ising model goes further, in that, depending on the number of states and on the dimensionality
of the system, also the Potts models exhibits a critical point which can be reached by tuning the temperature
\cite{1973JPhC....6L.445B}.
At the critical point the correlation length diverges and the properties of the system are governed by critical exponents.
The computation of critical exponents for the $q$-states Potts model at criticality goes almost as far back
as the same computation for the Ising and $O(N)$ models.

Critical exponents can be computed by means of the standard $\epsilon$-expansion in field theory \cite{Wilson:1973jj}.
There is a minor caveat here, because the analysis of a $S_q$ invariant theory requires the construction
of a Landau-Ginzburg $\phi^3$-like interaction (to be compared with $\phi^4$-like for the Ising and $O(N)$ models),
which in turn implies that the $\epsilon$-expansion should be constructed below the critical dimension $d_c=6$
\cite{Fisher:1978pf}, implying that the model has non-mean-field exponents also in $d=4,5$ \cite{1997PhLB..400..346B,2020arXiv200411289Z}.
This has the consequence that the $\epsilon$-expansion converges poorly, especially when one is interested
in estimates in lower dimensions such as $d=2,3$ \cite{1992JSP....67..553L,2011JPhA...44c2001D}.
For the same reason,
it is still very much unknown the precise form of the phase diagram of existence of
a non-trivial critical point for the Potts model in the $(q,d)$ plane \cite{Gorbenko:2018ncu,Gorbenko:2018dtm}.
Nevertheless the $\epsilon$-expansion offers a solid ground which is very reliable
when approaching the regions of small $\epsilon$ for obvious reasons,
so the method has been developed to the fourth order
\cite{Macfarlane:1974vp,deAlcantaraBonfim:1980pe,deAlcantaraBonfim:1981sy,Gracey:2015tta}.

Traditionally, critical points for $\phi^n$ theories with continuous and discrete symmetries, such as the ones mentioned above,
are mostly approached by tuning the temperature but maintaing the symmetry.
This can be conveniently implemented with functional perturbative renormalization group methods
\cite{ODwyer:2007brp,Osborn:2017ucf,Codello:2017hhh,Codello:2017qek,Codello:2018nbe}.
The exception to the previous statement is, of course,
the inclusion of a magnetic field which in the Ising and $O(N)$ models magnetizes the system,
while in the Potts model promotes one or more states
over the others.
A magnetic field breaks $S_q$ symmetry and drives the system
to an Ising-like $\mathbb{Z}_2$ phase \cite{Bonati:2010ce}.
In the field theory framework, the magnetic deformation of the system corresponds to
adding a linear term in the field(s) $h\cdot \phi$, having denoted $h$ the magnetic field.
There are, however, very many other ways to break explicitly the symmetry group of the system
other than including the magnetic field.
The simplest ones, which are also relevant in the sense of the renormalization group (RG),
are arbitrary non-singlet deformations which are quadratic and cubic in the field(s) $\phi$
\cite{Vasseur:2013baa,vjs,2017JPhA...50U4001C}.

The non-singlet deformations are important for several reasons. On the one hand, they can be
linked for some models to interesting observables, and therefore are worth studying
\cite{vjs,Vasseur:2013baa,2017JPhA...50U4001C,Gori:2017cyq}.
On the other hand, it is possible to imagine situations in which the temperature $T$ is tuned to its critical value $T_c$,
but the system is deformed by a non-singlet interaction in the Hamiltonian. In this case,
one can expect that the system exhibits critical behavior for this interaction going to zero,
which is in fact what happens \cite{wallace-young,Stephen:1977mw,Barbosa:1986kv,Theumann:1985qc}.
The new critical behavior is governed by so-called \emph{crossover exponents},
because non-singlet deformations drive the system to a phase with different symmetry,
and therefore at criticality the system crosses over two different phases.
A very simple example of critical crossover can be observed by giving a nonzero mass square $m^2$
to $k<N$ fields of th $O(N)$ model: the symmetry group is broken down to a subgroup
and if we take $m^2\to 0$ for $T\leq T_c$
there is a second order line separating $O(k)$ and $O(N-k)$ phases \cite{Calabrese:2002qi,DePrato:2003yd,Kompaniets:2019zes}.

The fact that non-singlet deformations are not scalars under the action of the group
does not imply that they have nothing to do with the group itself. Using representation theory,
it is possible to decompose any product of representations composing the general deformation
in terms of irreducible representations \cite{Vasseur:2013baa,vjs,2017JPhA...50U4001C},
even though, as we shall see later in section \ref{sect:deformations}, the application of this method
is complicated by the presence of mixing between vector currents.
One interesting fact is that, for almost all values
of the parameters $q$, the irreducibly decomposed operators are also scaling operators from the point of view of
the renormalization group, implying that the action of the dilatation operator commutes with the
action of the symmetry group. 
Even more interesting are the countable cases in which this commutation is not quite true:
in these cases the nondiagonalizable action of the dilatation is seen as a Jordan cell from the point of view of the underlying
conformal field theory (CFT) \cite{1984NuPhB.241..333B}, which leads to logarithmic contributions in the critical correlators
and logarithmic CFTs (log-CFTs)
\cite{Cardy:1999zp,2013JPhA...46W4001C,Vasseur:2013baa,Nivesvivat:2020gdj}
that have been observed numerically \cite{2014PhRvE..90d2106H,Tan2020,2019PhRvE..99e0103T,Ziff2020}
(see \cite{Hogervorst:2016itc} for a recent review, and see \cite{Zinati:2017hdy}
for an introductory discussion of log-CFTs from the point of view
of the renormalization group in the context of randomly diluted models).

One way to understand these logarithms goes as follows:
for some value $q=q_c$ the scaling dimensions of two or more operators
approach the same value and can be linearly combined into their would-be analog at $q_c$ and a logarithmic partner.
This combination happens precisely for those values $q_c$ in which the operators cannot be distinguished
on the basis of symmetry \cite{Vasseur:2013baa}.
%
%
For example, if we are considering the limit $q \to 1$ corresponding to the universality class of critical percolations,
clusters carry only one quantum number, therefore
the operator that measures (the correlations of) two distinct clusters of states, which we call the $2$-cluster operator,
has to somehow dissappear in the limit from general $q$.
In this case, the $2$-cluster operator becomes degenerate and combines with the singlet operator that measures the energy.
Even if the $2$-cluster operator is lost in the limit,
a new operator emerges from the combination which has logarithms in its two- and higher point functions \cite{vjs}.
The construction of observables that are actually capable
of evidencing the logarithmic behavior in the correlation is a whole new problem \cite{Tan2020,Ziff2020}.
Insofar, a well-crafted observable has been proposed and measured for the case $q\to 1$ \cite{vjs,Gori:2017cyq},
and is known to be relevant for the universality class to which (bond) percolations belong \cite{PhysRevB.41.9183,2000PhRvL..85.4104N}.
We argue, however, that similar observables can be constructed for arbitrary values of $q$
making room to several possible new numerical investigations of logarithmic critical behaviors \cite{Tan2020,Ziff2020}.

Having mentioned the case of the percolations, it is worth discussing some further point which are relevant
for few special values of $q$. Following literally the discussion on permutation invariance, the only values of $q$ allowed
are the integers bigger than or equal to two. Using a representation due to Fortuin-Kasteleyn \cite{Fortuin:1971dw},
it is however possible
to rewrite the partition function of the lattice Potts model in such a way that it makes sense
for arbitrary (continuous) values of $q$.
On a similar note, it is possible to show that the limit $q\to 1$ coincides with the model of bond percolations,
and thus at criticality covers the corresponding universality class \cite{PhysRevLett.35.327},
which remains a conformal field theory \cite{Gori:2015rta}.
Another interesting limit is the one for $q\to 0$, which is known to correspond to
a model of random clusters describing spanning trees and forests (depending on how the limit is taken)
\cite{Jacobsen:2003qp,Deng:2006ur,2007PhRvL..98c0602D}.
On the field theory side, the above two limits are not straightforwardly
accessible from the point of view of the degrees of freedom (they would correspond to
having either no fields, or a negative number). Even though there are mechanisms
to create the appropriate number of fields using fermions \cite{Caracciolo:2004hz}, and thus describe the correct physics,
the simplest option is to just compute any physical quantity (such as critical exponents) as a function
of $q$ and take the limits $q\to 0$ and $q\to 1$ at the end. In this sense, the limit $q\to 1$ of the field theory is known to
represent the universality class of percolations, while $q\to 0$ is believed to represent the universality class of spanning
forests and trees.

The interesting part of how clusters (including bond percolations, trees and forests) are generated through the Fortuin-Kasteleyn representation~\cite{Fortuin:1971dw}
is that they live in the links connecting the lattice sites. Thus, one can argue that the clusters live in
the same space in which propagator lines of the field theory would live. This connection becomes even more interesting
if complemented with the notion, recently discussed for the $O(N)$ model by Kompaniets and Wiese~\cite{Kompaniets:2019zes},
that the inverse of the scaling exponents of
a specific quadratic operator is the fractal dimension of propagator lines.
In the case of the $q$-states Potts model, representation theory shows that there are two other operators besides the singlet,
a vector and a tensor, which can both be associated to a certain notion of fractal dimension
\cite{1982JPhA...15.3829C,2003EPJB...34..479A,2010JSMTE..03..004A}.
The vector gives the fractal dimension of propagator lines just as in~\cite{Kompaniets:2019zes}, while
the tensor gives a fractal dimension of the cluster interpreted as a network of resistances \cite{harris-fisch,PhysRevB.18.416,dasgupta-et-al,2000AmJPh..68..896C}.

In this paper, we give estimates of non-singlet scaling exponents, crossover exponents,
and some further universal quantity relevant for log-CFT for the $q$-states Landau-Potts field theory
in $d=6-\epsilon$ using the renormalization group to three loops \cite{Osborn:2017ucf}.
Our result extends the previous literature \cite{wallace-young,Stephen:1977mw,Barbosa:1986kv,Theumann:1985qc} by
streamlining the computation, thanks to a careful use of the representation theory of $S_q$, and
including one further order in the perturbative expansion.
We also briefly discuss the same quantities for a sibling model
known as the hypertetrahedral model in $d=4-\epsilon$, again up to three loops,
which, to the best of our knowledge, were never presented elsewhere.
Since the hypertetrahedral model coincides with the well-known cubic model
(the model with the symmetry of a cube, $H_3$, \cite{aharony1973,carmona2000})
for $q=4$,
we provide numerical estimates of critical quantities using resummation methods in $d=3$.
Our results are used to compute several quantities of interest, but we pay particular attention to
crossover exponents and universal log-CFT data.
The paper is organized to give a brief (certainly inexhaustive) introduction to most of the topics that we touched
in this introduction. In Sect.~\ref{sect:potts-field-theory} we show how to construct general Landau-Ginzburg-like field
theories with permutation symmetry; in Sect.~\ref{sect:deformations} we discuss the irreducible representations
that emerge for quadratic and cubic deformations; in Sect.~\ref{sect:crossover-exponents} we discuss
how crossover exponents appear at criticality; in Sect.~\ref{sect:log-cft} we discuss the relation with log-CFT and several special limits;
in Sects.~\ref{sect:estimates-d6} and \ref{sect:estimates-d4} we give our main results for the cases $d=6-\epsilon$
and $d=4-\epsilon$ respectively.

\section{Landau-Potts field theories}\label{sect:potts-field-theory}

The construction of a field theory which is manifestly invariant under $S_q$ symmetry conveniently
starts by introducing a set $e^\alpha$ of $q$ vectors ($\alpha=1,\cdots,q$)
in $\mathbb{R}^N$ with $N=q-1$
forming the vertices of a regular $N$-symplex (hypertetrahedron) \cite{Zia:1975ha}.
The group $S_q$ then acts by permuting vertices of the symplex, while the vertices represents the possible \emph{states}
of the model.
For obvious reasons,
throughout this construction, we assume that $q$ takes values in $\mathbb{N}$ and $q>1$, although the most interesting
applications will come from analytic continuation outside this domain \cite{Fortuin:1971dw}.
We normalize the vertices as
\begin{equation}
\begin{split}
 &\sum_{i=1}^N e_i^\alpha e_i^\beta = q\delta^{\alpha\beta}-1 \,, \qquad
 \sum_{\alpha=1}^{q} e_i^\alpha = 0 \,, \\
 &\sum_{\alpha=1}^{q} e_i^\alpha e_j^\alpha = q\delta_{ij} \,.
 \label{Sqrel}
\end{split}
\end{equation}
Basic covariant tensors with respect to the permutation symmetry can be constructed as
sums of $n$ vertices
\begin{eqnarray}
 Q^{(n)}_{ij\cdots k} = \sum_{\alpha=1}^q e_i^\alpha e_j^\alpha \cdots e_k^\alpha\,,
 \label{Qdef}
\end{eqnarray}
which obviously implies that $Q^{(2)}_{ij}=q\delta_{ij}$ and $Q^{(1)}_i=0$,
but all other tensors are not as easily determined
for arbitrary values of $q$.
Since permutations act on the Greek indices $\alpha$, any sum over a Greek index is automatically invariant.

The general $S_q$ covariant order parameter is defined as the linear combination vector
$\psi^\alpha = \sum_{i=1}^{N} e^\alpha_i\varphi_i$ and therefore has $N$ independent components $\varphi_i$
which play the role of fields.
Notice that the covariance of $\psi$ under $S_q$ induces a corresponding
transformation on the multiplet $\varphi_i$, which we refer to as the vector representation of $S_q$.
Since the action on $\varphi_i$ is by construction an element of $O(N)$,
it underlies the fact that $S_{q}=S_{N+1}$ is a subgroup of
$O(N)$.\footnote{The group $O(N)$ is also the maximal symmetry
that a theory with this field content can have \cite{Osborn:2017ucf}.}
By construction, an arbitrary \emph{scalar} potential $V(\psi)$, which is a function of
$\psi^\alpha$,
is also invariant under the permutation group if it is summed over the state's index $\alpha$.
Consequently, inserting the explicit form of $\psi^\alpha$, it is always possible to express any potential
as the sum of products of the tensors $Q^{(n)}$ and the basic fields $\varphi_i$ and determine the explicit form of the action
of the permutation group.

In $d=6-\epsilon$ dimensions the perturbatively renormalizable Landau-Potts field theory has potential
\begin{eqnarray}\label{eq:cubic-potential}
 V(\psi) &=&  \frac{\lambda}{3!} \sum_\alpha (\psi^\alpha)^3
 =\frac{\lambda}{6} \sum_{ijk} Q^{(3)}_{i jk} \varphi_{i}\varphi_{j} \varphi_{k}\,,
\end{eqnarray}
which could be deformed by additional relevant operators that we analyze more carefully in the next section.
The cubic potential
ensures perturbative renormalizability in $d=6$ dimensions if complemented by the standard kinetic term
for the fields $\varphi_i$.
This can be seen from simple power-counting, which tells us that the coupling $\lambda$ is dimensionless in $d=6$.
Notice that, for $q=2$, it is easy to see that, by construction, $V(\psi)=0$:
this happens because the $2$-states Potts model is just a redefinition
of the Ising universality class, which in $d=6-\epsilon$ is above the upper critical dimension.

In $d=4-\epsilon$ dimensions the perturbatively renormalizable Landau-Potts field theory is referred
to as the \emph{hypertetrahedral model}
and sometimes as the \emph{restricted} Potts model. It can be seen as a generalization	
of the traditional $\phi^4$ theory of the Ising's universality class, which thus must be recovered for $q=2$ \cite{Zia:1975ha}.
The potential takes the form	
\begin{equation}\label{eq:quartic-potential}
\begin{split}
 V(\psi) &=	
 \frac{u}{4!}\frac{1}{q^2}\sum_{\alpha\beta} (\psi^\alpha)^2 (\psi^\beta)^2 + \frac{v}{4!}\sum_\alpha (\psi^\alpha)^4
 \\&=
 \frac{u}{24} \sum_{ij} (\varphi_{i})^2(\varphi_{j})^2	
 + \frac{v}{24} \sum_{ijmn} Q^{(4)}_{ijmn} \varphi_{i}\varphi_{j} \varphi_{m} \varphi_{n}\,.
\end{split}
\end{equation}	
Like in the previous example, we omitted any possible relevant terms and included only interactions	
which are power-counting marginal at the critical dimension. The first term of the potential is manifestly invariant	
under the larger group $O(N)$, which includes $S_q=S_{N+1}$ as subgroup given the identification $q=N+1$
with the number of states, and as such has been normalized to highlight this connection;	
in contrast the second term represents the departure from (internal) rotational invariance.	
The model with quartic interaction has the additional global reflection symmetry $\varphi_i \to -\varphi_i$,	
which combines with the permutation symmetry to form the group $S_q \times {\mathbb Z}_2$.	
The resulting symmetry group is such that	
for $N=3$ it coincides with the symmetry group $H_3$ of a cube
and therefore the analysis for $N=3$ must give the same results as the so-called cubic model.
We stress, however, that in general the hypercubic group does not coincide with the group of symmetries of the restricted model,
$H_N = ({\mathbb Z}_2)^N \rtimes S_N \neq S_{N+1}\times {\mathbb Z}_2$, so the two models are
distinct for any other finite number of components $N>3$.
The case $q=2$ (equivalently $N=1$) is special because the two interactions of \eqref{eq:quartic-potential}
coincide in the limit, so the model becomes the usual single-component $\phi^4$ of the Ising universality.

Interesting finite limits that can be followed through analytic continuation are $q\to 1$,
which is known for \eqref{eq:cubic-potential}
to be related to the universality class of percolations, and $q\to 0$, which is argued for \eqref{eq:cubic-potential}
to describe the universality class of special random cluster models representing spanning trees and forests
\cite{Jacobsen:2003qp,Deng:2006ur,2007PhRvL..98c0602D}.
Both identifications are supported by CFT analysis and by numerical simulations, to different extents.
Another interesting limit is $q\to \infty$, in which the spectrum of both Landau-Potts and hypertetrahedral models approach
$q$ non-interacting copies of the corresponding single-field limit \cite{Rong:2017cow}.
In $d=6-\epsilon$, this implies, for example, that the $S_q$
critical point of \eqref{eq:cubic-potential} becomes infinitely many independent Lee-Yang models with imaginary $\phi^3$ potentials \cite{Fisher:1978pf}.
The implication here is that a real fixed coupling (potential) for finite value $q$ must transition to an purely imaginary one for some value of $q$, and this happens at the value $q=10/3$, which is a well-known fact and can lead to further specifications on the nature of the $\epsilon$-series~\cite{1976JPhA....9.1441A}.
In $d=4-\epsilon$ the limit $q\to\infty$ leads to infinitely many copies of the spectrum of single-field's $\phi^4$.
Notice that the same happens when taking the limit $N\to\infty$ of the hypercubic model with $H_N$ symmetry~\cite{Rong:2017cow}.
A very qualitative way to interpret this limit would be to place the spins of the discrete model realizing the symmetry
(either the hypertetrahedron for $S_q$ or the hypercube for $H_N$) on a sphere of fixed radius, then take the limit of the number of embedding dimensions to infinity while keeping the radius fixed.
Then the spins become decoupled in the limit as the number of dimensions increase.

\section{Deformations of second and third order in the fields} \label{sect:deformations}

It is important to analyse all possible deformations of a critical theory  and we shall see how quadratic and cubic operators present different features.
In the presentation we follow in part Refs.~\cite{Vasseur:2013baa,2017JPhA...50U4001C}. 
The most general relevant deformation to the potentials \eqref{eq:cubic-potential} and \eqref{eq:quartic-potential}
with two copies of the field and no spatial derivatives is of the form $\phi^i\phi^j$ and involves an arbitrary symmetric source at the level of potential,
which has $\frac{N(N+1)}{2}=\frac{q(q-1)}{2}$ components.
Each copy of the field carries a standard representation of the permutation group, and  we can diagonalize the arbitrary quadratic deformation 
all in terms of the irreducible representations
\begin{equation} 
\begin{split}
 {\rm Sym}_2 (V \otimes V) = [q] \oplus [q-1,1] \oplus [q-2,2] \,.
\end{split}
\end{equation}
We denoted $S=[q]$ the singlet, $V= [q-1,1] $ the vector, and $T= [q-2,2]$ the symmetric $2$-tensor representations, whose dimensions sum up to $1+(q-1)+\frac{q(q-3)}{2}=\frac{q(q-1)}{2}$ exausting all possible symmetric tensors with two indices.

For almost all values of $q$, the terms of the above decomposition will also diagonalize the dilatations
and therefore be scaling directions.
The explicit forms of the quadratic operators are
\begin{equation} 
\begin{split}
 S^{(2)} = & \sum_i \varphi_i\varphi_i \,, \qquad \qquad
 V^{(2)}_k = \sum_{ij} \varphi_i \varphi_j Q^{(3)}_{ijk} \,, \\
 T^{(2)}_{mn} = &\sum_{ij} \varphi_i \varphi_j Q^{(4)}_{ijmn}
 -q(q-2)\varphi_m\varphi_n
 \\&
 -\frac{q}{q-1} \delta_{mn} \sum_i \varphi_i\varphi_i \,,
\end{split}
\end{equation}
in which we included an additional label $2$ in between brackets to denote that they are quadratic combinations in the number of fields~\cite{Vasseur:2013baa, Codello:2018nbe}.

Notice that the most general non-symmetric
decomposition would include an antisymmetric $2$-tensor, transforming under $A=[q-2,1,1]$.
Such tensor, however, does not contribute to the potential,
because it requires at least two spatial derivatives to construct it, $\varphi_{[i} \partial \varphi_{j]}$
(one if the operator is also allowed to carry spacetime spin).
This notion will become relevant also in the forthcoming analysis of the cubic sector.

The most general deformation of the potential involving three copies of the field and no spatial derivatives
is of the form $\varphi_i\varphi_j\varphi_k$ and can be dealt with
in a similar fashion. Using the same assumptions as the previous case,
the irreducible decomposition of the product of three vector representations is
\begin{equation} \label{eq:cubic-irreps}
\begin{split}
 {\rm Sym}_3 (V \otimes V \otimes V) = & [q] \oplus 2 \, [q-1,1] \oplus [q-2,2] 
 \\&
 \oplus [q-2,1,1] \oplus [q-3,3] \,,
\end{split}
\end{equation}
corresponding to the following dimensions
\begin{equation}  
\begin{split}
\label{dim-decomp}
\frac{(q-1)q(q + 1)}{6} &= 1 + 2 (q - 1) + \frac{q(q - 3)}{2}
\\&
+ \frac{(q - 1)(q - 2)}{2}
+ \frac{q(q - 1)(q - 5)}{6}\,.
\end{split}
\end{equation}
If compared with the previous example, the decomposition contains also
the symmetric $3$-tensor representation $Z=[q-3,3]$.
As before we neglect the antisymmetric contribution because
it is not realized by our cubic deformations \cite{Vasseur:2013baa,2017JPhA...50U4001C}.
A partial list of the cubic operators has been discussed in~\cite{Vasseur:2013baa}, and, in our notation, it is given by
\begin{equation}\label{eq:cubic-ops-1}
\begin{split}
 & S^{(3)} = \sum_{ijk} \varphi_i \varphi_j \varphi_k Q^{(3)}_{ijk}\,, 
 \\&
 V^{(3)}_k = \sum_{ijm} \varphi_i \varphi_j \varphi_m Q^{(4)}_{ijmk}
 -\frac{1}{2} \varphi_k \sum_i \varphi_i\varphi_i
 \\&
 T^{(3)}_{mn} =\sum_{ijk} \varphi_i \varphi_j \varphi_k Q^{(5)}_{ijkmn}
 - (q-2) \sum_{ij} \varphi_i \varphi_j \varphi_{(m} Q^{(3)}_{n)ij}
\\&
 \quad- \sum_{ij} \varphi_i\varphi_i \varphi_j Q^{(3)}_{mnj}
 -\frac{1}{q-1} \delta_{mn} \sum_{ijk} \varphi_i \varphi_j \varphi_k Q^{(3)}_{ijk} 
\end{split}
\end{equation}
for first three representations, and
\begin{equation}\label{eq:cubic-ops-2}
\begin{split}
 & Z^{(3)}_{mnp} = \sum_{ijk} \varphi_i \varphi_j \varphi_k Q^{(6)}_{ijkmnp} 
 \\
 \quad & -  \frac{(q-2)(q-3)}{2(q-1)}\sum_{rs} \varphi_r \varphi_s \varphi_m Q^{(4)}_{nprs}  +{\rm [2 ~ perms]}
 \\ 
 \quad & -\frac{1}{q-1} \sum_{rst} \delta_{mn} \varphi_r \varphi_s \varphi_t Q^{(4)}_{prst} +{\rm [2 ~ perms]}
 \\
 \quad & -\frac{q-3}{q(q-1)} \sum_{rst} \varphi_r \varphi_s \varphi_t Q^{(3)}_{mnr} Q^{(3)}_{pst} +{\rm [2 ~ perms]}
 \\
 \quad & -\frac{2}{q(q-1)(q-2)} \sum_{rst} \varphi_r \varphi_s \varphi_t Q^{(3)}_{rst} Q^{(3)}_{mnp}
 \\
 \quad & - \frac{3}{q-1} \varphi^2 \sum_{r} \varphi_r Q^{(4)}_{mnpr}
 +\frac{q^2(q-3)(q-4)}{2(q-1)} \varphi_m \varphi_n \varphi_p
 \\
 \quad & +\frac{q^2}{2(q-1)} \delta_{mn} \varphi^2 \varphi_p +{\rm [2 ~ perms]}
\,,
\end{split}
\end{equation}
for the new one, where $\varphi^2=\sum_{r} \varphi_r \varphi_r$,
``perms" refers to two more cyclic permutations of the indices $(mnp)$, and
the new label $3$ reminds us that there are three copies of the fields.
The general construction comes by iteratively removing the traces among all indices~\cite{Vasseur:2013baa},
but let us stress that it is not complete, as further discussed in~\cite{2017JPhA...50U4001C}.
Moreover, there can be operators which do not transform according to any representation of $S_q$,
as we discuss in Sect.~\ref{sect:example} with an explicit example.

First of all, we realize that in \eqref{eq:cubic-irreps} there are actually two vectors
whose subspace is generated by $\sum_{ijm} \varphi_i \varphi_j \varphi_m Q^{(4)}_{ijmk}$
and $\varphi_k \sum_i \varphi_i\varphi_i$,
so that a mixing altering the form of $V^{(3)}_k$ of Eq.~\eqref{eq:cubic-ops-1} is allowed.
We therefore modify $V^{(3)}_k$ to include a $2$-dimensional vector space (modulo the overall normalization)
by introducing an arbitrary constant $B\neq1$
\begin{equation}\label{eq:cubic-vector-ops}
\begin{split}
 V^{(3)}_k = \sum_{ijm} \varphi_i \varphi_j \varphi_m Q^{(4)}_{ijmk}
 -\frac{B}{2} \varphi_k \sum_i \varphi_i\varphi_i \,.
\end{split}
\end{equation}
Since the vector is constructed from two monomials, the computation of the scaling operator
requires the diagonalization of a two-by-two matrix that depends on $B$ and $q$.
This results in an equation with two solutions, $B=B(q)$ and $B=B'(q)$,
coming from a quadratic algebraic equation for the scaling deformations.

For the Landau-Potts model the vector operators must be treated with additional care when comparing RG and CFT results.
From the point of view of CFT, the operator $V^{(2)}_k$ combines with the descendant $\frac{\partial V}{\partial\phi_i}$ through the equations of motion of \eqref{eq:cubic-potential}, and therefore it should have scaling dimension equal to the one of the field plus two, $\Delta_{V^{(2)}} = 2+\Delta_\phi$. This scaling relation is replaced, in the RG context, by the more general relation
involving the critical exponents $\theta_\phi+\theta_{V^{(2)}}=d$, which we verify explicitly.
For an arbitrary primary operator ${\cal O}$ the relation between scaling dimensions and critical exponents is of course $\Delta_{{\cal O}}=d-\theta_{{\cal O}}$, but it is modified for descendants.
Using the relation $\Delta_\phi=(d-2+\eta)/2$ and the above two formulas,
we find that the standard relation is modified by the presence of the anomalous dimension,
$\Delta_{V^{(2)}}=d-\theta_{V^{(2)}}+\eta$.

From the physical point of view, two irreducible representations of the same rank but of different order (such as $S^{(2)}$ and $S^{(3)}$)
share the same $S_q$ quantum number, but have different scaling dimensions. They can be thought to
contribute different corrections to the same quantity from the point of view of statistical mechanics.
For example, if we associate the quantum numbers of the singlet to the energy, then the leading nontrivial correction comes from the scaling operator $S^{(2)}$, and the subleading correction from $S^{(3)}$.\footnote{In this language,
the leading correction to the energy is actually the identity operator $S^{(0)}=1$, which has trivial scaling dimension $d$
and therefore is not particularly interesting to report.}

\section{Crossover exponents and fractal dimensions} \label{sect:crossover-exponents}

There are two main reasons to discuss the scaling behavior of the most general quadratic and cubic operators.
The first one is that, with the exception of the singlet operators, they all break the underlying permutation symmetry of
the Potts field theory, and therefore they allow us to describe the \emph{crossover} of the system from the symmetric critical
point to a symmetry-broken phase. Under the assumption that the symmetry-broken phase has some residual (unbroken) symmetry, this implies that, close to the critical point, some crossovers to phases with smaller symmetries have associated crossover exponents.
To clarify this statement, first assume that the critical point is described by a certain CFT action $S_*$ and that, within a symmetric regime,
we reach criticality
by tuning the energy operator $E$ as in $S=S_*+g E$. We know that $g \sim t^{\frac{1}{\nu}} $ which defines the critical exponent $\nu$, following the usual scaling assumption that $t$ is the reduced temperature $t\sim (T-T_c)/T_c$. Now perturb $S_*$ by a further operator $\tilde{E}$ which breaks the symmetry in some way, $S=S_*+g E + g_1 \tilde{E}$, and which belongs to some scaling multiplet, implying the behavior $g_1 \sim t^{\tilde{\theta}} $ close to the critical point.
Solving the two scaling relations in $t$, we can derive how the symmetry breaking part scales with the symmetric coupling close to criticality
$g_1 \sim g^{\Phi} $, which defines the crossover exponent $\Phi=\tilde{\theta} \nu$ of $\tilde{E}$.

Given a symmetry-breaking operator $\tilde{E}$ of interest, the only remaining point is to establish whether
it actually belongs to
a scaling multiplet or not. At the quadratic level the interesting operators have either one or two representation indices,
implying that they can either realize a breaking in which either one or two components are selected.
In the first case, corresponding to the vector $V^{(2)}_k$ the crossover exponent is determined by a scaling relation to be equal to the thermodynamical exponent $\beta$ when $d=6-\epsilon$.
In the second case we have a completely nontrivial breaking $S_q \to \mathbb{Z}_2$, in which the Potts model descends into an Ising phase \cite{Bonati:2010ce}.

The critical exponents of quadratic deformations are also interesting from a more geometrical perspective.
It has been shown in~\cite{Kompaniets:2019zes}
that the renormalization of a specific quadratic insertion of the form
${\cal O} \sim \phi^2_i -\phi_j^2$ with $i\neq j$ {(equivalently $\phi_i \phi_j$ in a rotated basis)}
allows for the determination of the fractal dimension $d_f$ of propagator lines as $d_f=\theta$,
where we defined $\theta$ the critical exponent of the operator that contains ${\cal O}$ in its multiplet.
Direct inspection of the irreducible operators for some low values of $q$
shows that the above operator is found in the irreducible quadratic vector, implying that the fractal dimension
of propagator lines can be estimated direcly from the critical exponent of $V^{(2)}_k$. In $d=6-\epsilon$ the same exponent is subject to a
scaling relation, which thus relates $d_f$ and the thermodynamical exponent $\beta$ as we shall see later.
This geometrical notion is particularly important when the lattice model is a cluster model that
lives in the space of the links, rather than that of the sites,
because the fractal dimension of the propagator lines becomes a fully fledged fractal dimension of
the clusters. Therefore, our analysis is relevant to estimate the fractal dimensions, for example,
of percolation's clusters, which are known to belong to the universality class $q\to 1$,
and of clusters of spanning forests and trees, which are believed to belong to the universality class $q\to 0$.
Furthermore, the scaling of the tensor deformation has been linked to the conductivity behavior of the underlying cluster,
which can be thought as a fractal dimension $d_r$ in itself~\cite{Theumann-Gusmao1984}. We report both fractal dimensions in the next sections.

\section{Relation with log-CFTs} \label{sect:log-cft}

The second reason why we include all possible deformations has to do with the construction of logarithmic CFTs.
To understand where logarithms are coming from, let us first recall that the expressions of the operators for arbitrary values of $q$ (the number of states) are certainly redundant if $q$ is specialized to some small values such as, for example, $q_c=1,2,3$ etc.
This redundancy implies that, by taking $q$ equal to $q_c$, two or more operators of Sect.~\ref{sect:deformations} and of the same rank will degenerate to the same multiplet and the spectrum of the theory will encounter some form of discontinuity.
The usefulness of the general approach is that we can analitically continue $q\to q_c$,
study the onset of the degeneracy, and discuss the physical implications of the two operators colliding.
Rather generally, it has been shown that, given two operators $X$ and $\tilde{X}$ with scaling dimensions
$\Delta_X(q)$ and $\Delta_{\tilde{X}}(q)$ coinciding for some value $q=q_c$ ($\Delta_c\equiv \Delta_X(q_c)=\Delta_{\tilde{X}}(q_c)$),
in the limit $q\to q_c$ the action of the dilatation operator is not appropriately diagonalized because of the appearance of
a Jordan cell \cite{2013JPhA...46W4001C,Hogervorst:2016itc,2019PhRvE..99e0103T}.
The two operators are bundled in a so-called logarithmic pair in which the usual powerlaw behavior 
of CFT makes room for a logarithmic term and a new universal quantity defined as
$$
 \alpha_X = \lim_{q \to q_c} \frac{\Delta_X-\Delta_{\tilde{X}}}{q-q_c}\,.
$$
A clarifying example is certainly in order, so we provide one in the rest of this section.

The simplest and, arguably, most interesting example of this situation happens for $q\to q_c=1$, which corresponds to the universality class of critical percolations. 
We follow~\cite{vjs}, albeit with some minor notational differences, to provide an example of logarithmic pair.
The role of the above operators $X$ and $\tilde{X}$ is taken by the energy singlet $E$, which has two point function 
\begin{equation}
 \begin{split}
 \langle E(x) E(0) \rangle = (q-1) \frac{ A(q) }{\left|x\right|^{2\Delta_E(q)}}\,,
 \end{split}
\end{equation}
and by the $2$-cluster operator $\tilde{E}$, which has two point function
\begin{equation}
 \begin{split}
 &\langle \tilde{E}_{ij}(x) \tilde{E}_{kl}(0) \rangle =  \frac{2}{q^2}
  \Bigl(\delta_{ik}\delta_{jl}+\delta_{ik}\delta_{jl}
 \\& \quad -\frac{\delta_{ik}+\delta_{il}+\delta_{jk}+\delta_{jl}}{q-2}
  +\frac{2}{(q-1)(q-2)}\Bigr)
 \frac{\tilde{A}(q)}{\left|x\right|^{2\Delta_{\tilde{E}}}}\,,
 \end{split}
\end{equation}
respectively.
The normalizations $A(q)$ and $\tilde{A}(q)$ are regular functions in the vicinity of $q=1$, 
and it is therefore evident that one correlator vanishes while the other diverges in the limit $q\to1$. 
This singular behavior happens regardless when normalizing correlators for general $S_q$ symmetry 
and it is intimately connected with the appearance of logarithms that we are about to see.

In the limit $q\to 1$, we have that the scaling dimensions of the two operators coincide, so we denote them $\Delta_E(1)=\Delta_{\tilde{E}}(1)\equiv\Delta_c$ as before.
A physical way to understand why something like this should happen comes
from realizing that for $q=1$ there cannot be an independent $2$-cluster operator,
because there are no two differently labelled clusters: there is only one which is the one of the percolation itself.
Consistency with the two operators becoming degenerate
requires that also the normalizations are related, which in this case implies that only one quantity survives
the limit $A(1)= \tilde{A}(1)\equiv A$.
Let us define the universal quantity
\begin{equation}\label{eq:alpha-universal}
 \begin{split}
\alpha_E \equiv \lim_{q\to 1} \frac{\Delta_{\tilde{E}}-\Delta_E}{q-1}\,,
 \end{split}
\end{equation}
in which we specialized the previous definition to the case $q_c=1$.
We define a new operator from the combination
\begin{equation}
 \begin{split}
 \hat{E}_{ij}(x) \equiv \tilde{E}_{ij}(x) +\frac{2}{q(q-1)} E(x)\,.
 \end{split}
\end{equation}
The new operator is meaningful in the limit $q\to 1$ because the singular behavior of the two terms cancels out.
To see this, one inspects the correlator and takes the limit $q\to 1$;
using all the above formulas one finds
\begin{equation}
 \begin{split}
 &\langle \hat{E}_{ij}(x) \hat{E}_{kl}(0) \rangle = 2
  \Bigl(\delta_{ik}\delta_{jl}+\delta_{ik}\delta_{jl} 
 \\& \quad
 +\delta_{ik}+\delta_{il}+\delta_{jk}+\delta_{jl}
  +4 \alpha_E \log \left|x\right| \Bigr)
 \frac{A}{\left|x\right|^{2\Delta}}\,.
 \end{split}
\end{equation}
A very interesting aspect of this construction is that there are special observables which allow to measure the logarithmic corrections. For example, in the case of percolations, one can observe with a Monte Carlo measurement
the logarithmic behavior and verify explicitly  a dependence in agreement
with the leading order analytic computation for
$\alpha_E=\frac{\sqrt{3}}{\pi}\simeq 0.5513$.

Contact with the operators shown in Sect.~\ref{sect:deformations} is made
using the identificatios $E=S$ and $\tilde{E}=T$,
since they carry the same quantum numbers \cite{2013JPhA...46W4001C}.
This tells us that, for the case of percolations $q=1$, we can compute the universal quantity \eqref{eq:alpha-universal}
for both the leading and subleading nontrivial contributions to the energy (respectively $S^{(2)}$ and $S^{(3)}$),
thus resulting in the coefficients $\alpha_S^{(2)}$ and $\alpha_S^{(3)}$, that will be shown in the next section.
In this case, it is the leading coefficient that should be compared with the lattice measure,
although for obvious reasons the $\epsilon$-expansion converges poorly.
For different values of $q$, there will be different Jordan cells coming from difference operator degeneracies
\cite{2013JPhA...46W4001C},
which will also be discussed in the next sections.

\section{RG estimates from $d=6-\epsilon$} \label{sect:estimates-d6}

The renormalization group for the potential \eqref{eq:cubic-potential} can be found up to the third loop order
in~\cite{Osborn:2017ucf} in a dimensionally regulated scheme with minimal subtraction in $d=6-\epsilon$.
We make use of a general theoretical framework which we call functional perturbative RG~\cite{ODwyer:2007brp, Codello:2017hhh} 
and which is linked to a leading order CFT approach~\cite{Codello:2019vtg} which was considered recently~\cite{Rychkov:2015naa, Codello:2017qek, Codello:2018nbe}.
All our results can be extracted making use of the flow of the dimensionless renormalized potential
$v(\varphi)= \mu^{-d} V(\mu^{d/2-1}Z\cdot\varphi)$, which includes the RG scale $\mu$ as well as a wavefunction renormalization matrix $Z_{ij}$ leading to a matrix of anomalous dimensions $\gamma=-\mu\partial_\mu \log Z$.
For illustrative purpose we display the one loop flow of \eqref{eq:cubic-potential}
\begin{eqnarray} \label{eq:beta-potential-d6}
\begin{split}
 \beta_v &= -d v + \frac{d-2}{2}  v_i \varphi_i +  v_i \gamma_{ij} \varphi_i -\frac{1}{6}  v_{ij}v_{jk}v_{ki}\,, \\
 \gamma_{ij} &= \frac{1}{2}v_{ikl}v_{jkl}\,,
\end{split}
\end{eqnarray}
in which labels on $v(\varphi)$ correspond to derivatives with respect to the components $\varphi$
and repeated indices are summed over. 
As said it can be obtained both with RG and CFT methods and some general analysis can be found in~\cite{Osborn:2017ucf, Codello:2019isr}.
An advantage of the approach is that all possible quantities related to the scaling  of relevant and marginal spinless operators can be derived from the above equation. For the beta function of the coupling $\lambda$ as in \eqref{eq:cubic-potential}, it is sufficient to insert the explicit form of the potential in \eqref{eq:beta-potential-d6} and simplify the result using the definitions of the hypertetrahedron vertices~\eqref{Qdef} and Eq.~\eqref{Sqrel}.

Composite and relevant operators, such as those of the previous sections,
are similarly renormalized by deforming and linearizing the flow $v(\varphi)\to v(\varphi)+{\cal O}(\varphi)$,
where ${\cal O}(\varphi)$ is the operator in question.
In practice, we have that ${\cal O}(\varphi)$
is spinless\footnote{Here and in the following, ``spinless'' refers to field's space (internal) indices and not spacetime's and
it means that the operator does not have uncontracted field indices.
} 
and, therefore, we have to turn on each operator belonging to the irreducible representations
by introducing an oppurtune source
\begin{equation}
\begin{split}
 {\cal O}(\varphi) = \sum_{X=S,V,T,Z,\cdots} J_X \cdot X(\varphi)\,,
\end{split}
\end{equation}
which can be done for both quadratic and cubic deformations. The renormalization of the sources
$J_X$ is multiplicative and easily leads to the corresponding critical exponents $\theta_X$ as a function of
the couplings.

At the critical point the anomalous dimension matrix is diagonalized to $\gamma_{ij}=\frac{\eta}{2}\delta_{ij}$ defining the anomalous dimension, while the fixed point coupling 
$\lambda^*$ admits an expansion in $\sqrt{\epsilon}$ coming from the solution of its beta function.
The expression of the critical coupling is used inside all RG function to determine the critical exponents in the $\epsilon$-expansion.


We denote with $\theta^{(n)}_X$ the critical exponent associated with the relevant deformation containing $n$ fields and irreducible multiplet $X=S,V,T,Z$ (which is enough up to $n=3$). Starting with the bilinear operators, we find in the $\epsilon$-expansion  (for results at second order see~\cite{Theumann-Gusmao1984})
\begin{widetext}
\begin{eqnarray} \label{eq:quad-deformations-d6}
 \theta^{(2)}_S &=& 2
 -\frac{5 (q-2) \epsilon}{3 (3 q-10)}
 -\frac{(q-2) \left(43 q^2-290 q+900\right) \epsilon ^2}{54 (3 q-10)^3}
 +\epsilon ^3 \Bigl\{\frac{4(q-2) \left(2 q^3-3 q^2+32 q-120\right) \zeta_3}{3 (3 q-10)^4}
 \\&&
 -\frac{(q-2) \left(8375 q^4-91790 q^3+330344 q^2-537720 q+622800\right)}{1944 (3 q-10)^5}\Bigr\}
 \nonumber 
\end{eqnarray}
\begin{eqnarray}
 \label{eq:quad-deformations-d6-2}
 \theta^{(2)}_V &=& 2
 -\frac{(5 q-16) \epsilon }{3 (3q-10)}
 -\frac{(q-2) \left(43 q^2-257 q+420\right) \epsilon ^2}{54 (3 q-10)^3}
 +\epsilon ^3 \Bigl\{\frac{8(q-3) (q-2) \left(q^2-6 q+13\right) \zeta_3}{3 (3 q-10)^4}
 \\&&
 -\frac{(q-2) \left(8375 q^4-101525q^3+464504 q^2-958836 q+774720\right)}{1944 (3 q-10)^5}\Bigr\}
 \nonumber 
\end{eqnarray}
\begin{eqnarray}
 \label{eq:quad-deformations-d6-3}
 \theta^{(2)}_T &=& 2
 +\frac{(q+4) \epsilon }{3 (3q-10)}
 +\frac{(q-2) \left(86 q^2+101 q-840\right) \epsilon ^2}{54 (3 q-10)^3}
 +\epsilon ^3 \Bigl\{
 -\frac{4 (q-2) \left(4 q^3-3 q^2-74 q+162\right) \zeta_3}{3 (3 q-10)^4}
 \\&&
 +\frac{(q-2)\left(16750 q^4-67603 q^3-233960 q^2+1384884 q-1432080\right)}{1944 (3 q-10)^5}
 \Bigr\}\,.
 \nonumber 
\end{eqnarray}
%
The critical exponent $\theta^{(2)}_S$ is related to the scaling of the correlation length by $\nu= 1/\theta^{(2)}_S$.
Following the discussion of the previous sections, the exponents $\theta^{(2)}_V$ and $\theta^{(2)}_T$
have the meaning of fractal dimension of the clusters $d_f=\theta^{(2)}_V$,
and of the conductivity which we denote $d_{r}=\theta^{(2)}_T$.
We verify in general the scaling relation $\theta^{(2)}_V=(d-2+\eta)/2$
by Wallace and Young~\cite{wallace-young}, which implies $d_f=(d-2+\eta)/2$.
For completeness we give the anomalous dimension too
\begin{eqnarray}
 \label{eq:eta-d6}
 \eta &=& 
 -\frac{(q-2) \epsilon }{3 (3 q-10)}
 -\frac{(q-2) \left(43 q^2-257 q+420\right) \epsilon ^2}{27 (3 q-10)^3}
 +\epsilon ^3 \Bigl\{\frac{16 (q-3) (q-2) \left(q^2-6 q+13\right) \zeta_3}{3 (3q-10)^4}
\\&&
 -\frac{(q-2) \left(8375 q^4-101525 q^3+464504 q^2-958836 q+774720\right)}{972 (3 q-10)^5}\Bigr\}
\,.
 \nonumber 
\end{eqnarray}

The conductivity fractal dimension is most interesting for clusters with a simple geometric interpretation, which are $q=1$ (percolations) and $q=0$ (spanning trees and forest). In these limits we find
\begin{eqnarray}
 d_{r,\, {\rm s.f.}} &=&
 2-\frac{2 \epsilon }{15}-\frac{7 \epsilon^2}{225}
 +\left(\frac{27 \zeta_3}{625}-\frac{221}{15000}\right) \epsilon ^3 \,,
 \\
 d_{r,\, {\rm perc.}} &=&
 2-\frac{5 \epsilon }{21}-\frac{653 \epsilon^2}{18522}
 +\left(\frac{356 \zeta_3}{7203}-\frac{332009}{32672808}\right) \epsilon ^3 \nonumber\,.
\end{eqnarray}

The crossover exponents are defined as $\overline{\Phi} = \nu \theta^{(2)}_V$ and $\Phi= \nu \theta^{(2)}_T$, so that they
govern the scaling close to criticality of non-symmetric deformations.
We find
\begin{eqnarray}
 \overline{\Phi} &=& 
 1+ \frac{\epsilon }{3 q-10}
 +\frac{(q-2) (79 q-140) \epsilon ^2}{36 (3 q-10)^3}
 +\epsilon ^3 \Bigl\{
  -\frac{2 (q-2) (5 q^2-10 q-14) \zeta_3}{(3 q-10)^4}
 \nonumber \\&&
 +\frac{(q-2) (11903 q^3-110840 q^2+322172 q-242640)}{1296 (3 q-10)^5}
 \Bigr\} 
 \\
 \Phi &=&
 1+\frac{(q-1) \epsilon }{3 q-10}
 +\frac{(q-2) (q-1) (133 q-320) \epsilon ^2}{36 (3 q-10)^3}
 +\epsilon ^3 \Bigl\{
  -\frac{4 (q-2) (q-1) (q^2-7) \zeta_3}{(3 q-10)^4}
 \nonumber \\&&
 +\frac{(q-2) (q-1) (21893 q^3-152996 q^2+287972 q-30240)}{1296 (3q-10)^5}
 \Bigr\} 
\end{eqnarray}
The first crossover exponent is constrained by the usual scaling relation
to be equal to the thermodynamical exponent $\beta=\nu(d-2+\eta)/2=\overline{\Phi}$.
The second exponent is instead trivial for the percolations $q\to 1$, becoming $\Phi=1$:
this limit we verify at any order in the expansion, but can also be shown in general.

The critical exponents of deformations with three copies of the fields are
\begin{eqnarray}
 \theta^{(3)}_S &=&
 \epsilon
 -\frac{\left(125 q^2-794 q+1340\right) \epsilon ^2}{18 (3 q-10)^2}
 +\epsilon ^3 \Bigl\{\frac{4\left(5 q^3-44 q^2+158 q-212\right) \zeta_3}{(3 q-10)^3}
 \\&&
 +\frac{36755 q^4-466622 q^3+2303256q^2-5184856 q+4408720}{648 (3 q-10)^4}\Bigr\}
 \nonumber \\
 \theta^{(3)}_T  &=&
 -\frac{(q+20)\epsilon }{3 (3 q-10)}
 -\frac{\left(86 q^3-627 q^2+8250 q-21800\right) \epsilon ^2}{54 (3 q-10)^3}
 +\epsilon ^3 \Bigl\{\frac{4 \left(4 q^4+5 q^3+104 q^2-1520 q+3360\right) \zeta_3}{3 (3 q-10)^4}  \label{theta3T}
 \\&&
 -\frac{16750 q^5-125003 q^4+72906 q^3+6185540 q^2-33921480 q+50813600}{1944 (3 q-10)^5}\Bigr\}
 \nonumber \\
 \theta^{(3)}_Z  &=&
 -\frac{2 q \epsilon }{3 q-10}
 -\frac{q \left(43 q^2+272 q-1220\right) \epsilon ^2}{18 (3 q-10)^3}
 +\epsilon ^3 \Bigl\{\frac{4 q \left(2 q^3+11 q^2-142 q+288\right) \zeta_3}{(3 q-10)^4}\label{theta3Z}
 \\&&
 -\frac{q \left(8375 q^4+17172 q^3-391224 q^2+917456 q-327120\right)}{648 (3 q-10)^5}\Bigr\} \,.
 \nonumber 
\end{eqnarray} 
\end{widetext}
All complete beta functions and critical exponents are provided through an ancillary file.

We do not give $\theta^{(3)}_V $ and $\theta^{(3)\prime}_V$ for arbitrary $q$ because for the vector subsector, which
requires the diagonalization of a two dimensional subspace, the final result is given by expressions very involved due to radicals.
Instead, we give them only for the first few interesting values $q=0,1,2,3$.
For spanning forests
\begin{equation}
\begin{split}
  \theta^{(3)}_V|_{q=0} &= \frac{17 \epsilon }{15}
 -\frac{206 \epsilon^2}{225}
 +\left(\frac{633 \zeta_3}{625}+\frac{377527}{405000}\right) \epsilon ^3\,,
 \\
 \theta^{(3)\prime}_V|_{q=0} &=0 \,, 
\end{split}
\end{equation} 
for percolations
\begin{equation}
\begin{split}
&
\left.\begin{matrix}
\theta^{(3)}_V|_{q=1} \\
\theta^{(3)\prime}_V|_{q=1}
\end{matrix}\right\}
=
  \frac{25\pm\sqrt{889}}{42}  \epsilon
    -\frac{2664841 \pm 92599 \sqrt{889} }{4704588}\epsilon^2
 \\&\qquad
 +\Bigl\{\frac{50 \left(4699\pm165 \sqrt{889}\right) \zeta_3}{304927}
 \\&\qquad
  +\frac{11 \left(21667198601\pm725975407\sqrt{889}\right)}{351319813488}\Bigr\} \epsilon^3
\,,
\end{split}
\end{equation}
for $q=2$
\begin{equation}
\begin{split}
  \theta^{(3)}_V|_{q=2}& =
  \frac{11 \epsilon}{6}
  -\frac{445 \epsilon ^2}{216}
 +\left(\frac{35 \zeta_3}{8}+\frac{53375}{15552}\right)
   \epsilon ^3
\,, \\
 \theta^{(3)\prime}_V|_{q=2} &=
 -\frac{\epsilon}{2}
\,, 
\end{split}
\end{equation}
and finally for $q=3$
\begin{equation}
\begin{split}
 \theta^{(3)}_V|_{q=3}  &= 6 \epsilon
 -\frac{17 \epsilon ^2}{6}
 +\left(180 \zeta_3+\frac{1713}{8}\right) \epsilon ^3 \,, 
 \\
 \theta^{(3)\prime}_V|_{q=3} &= -\frac{11 \epsilon}{3}
 -\frac{158 \epsilon ^2}{9}
 -\left(16 \zeta_3+\frac{17380}{81}\right) \epsilon ^3\,.
\end{split}
\end{equation}

\begin{figure}[htb]
  \includegraphics[width=0.49\textwidth]{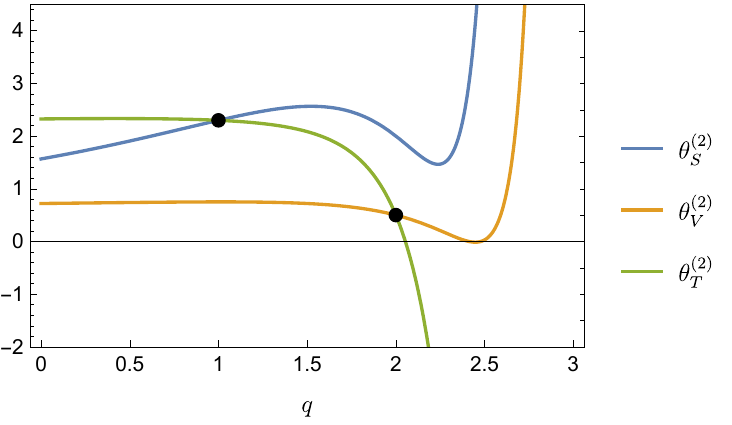}
  \caption{\label{fig:theta2s-d6} Plot of the critical exponents $\theta^{(2)}_X$ for $X=S,V,T$ of the Landau-Potts model
  extrapolated to $d=3$ ($\epsilon=3$) as a function of $q$.
  The marked points highlight the degeneracies that lead to logarithmic behavior.}
\end{figure}

\begin{figure}[htb]
  \includegraphics[width=0.49\textwidth]{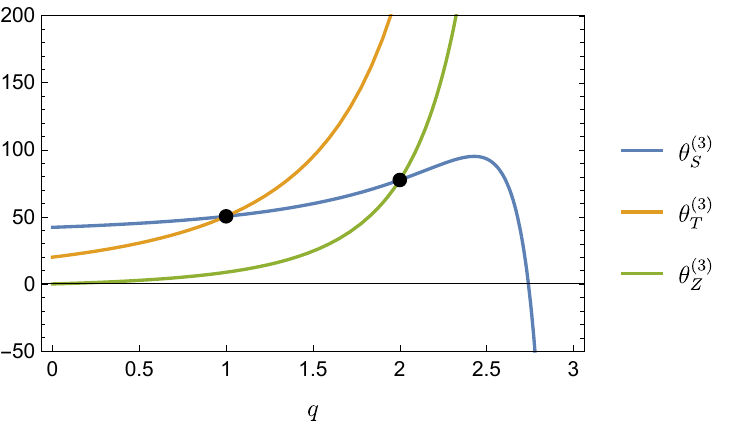}
  \caption{\label{fig:theta3s-d6} Plot of the critical exponents $\theta^{(3)}_X$ for $X=S,T,Z$ of the Landau-Potts model
  extrapolated to $\epsilon=3$ as a function of $q$. The marked points highlight some of the degeneracies that lead to logarithmic behavior.}
\end{figure}

There are several predictable patterns of degeneracy that we can check as a function of $q$. A simple explanation goes as follows:
because of their quantum numbers, the $n$-tensor operators are interpreted as generators of $n$-cluster functions,
but, e.g.~if $n>q$, there are not enough states to label all different clusters, implying that the $n$-cluster operator must degenerate with some lower cluster.
The pattern of degeneracies has been carefully explained in~\cite{Vasseur:2013baa,vjs,2017JPhA...50U4001C}, here we just summarize our explicit findings
based on those expectations.
\begin{itemize}
 \item For $q=1$, corresponding to the universality class of percolations, the energy and $2$-cluster operators degenerate,
 which implies a relation for the leading and subleading scaling exponents
 $\theta^{(M)}_S|_{q=1}=\theta^{(M)}_T|_{q=1}$ with $M=2,3$.
 This is a well-known relation that has implications for the logarithmic behavior of percolations
 and allows to construct a logarithmic observable. 
 \item For $q=2$, corresponding to the Ising universality class, the subleading energy and $3$-cluster operators degenerate,
 but also the subleading magnetization and the $2$-cluster operators degenerate.
 We have therefore
 $\theta^{(3)}_S|_{q=2}=\theta^{(3)}_Z|_{q=2}$ and
 $\theta^{(M)}_V|_{q=2}=\theta^{(M)}_T|_{q=2}$ for $M=2,3$.
 However, upon inspection these relations become rather trivial because this is the Ising universality class
 in $d=6$ is above its upper critical dimension and therefore here has mean-field critical exponents.
 \item For $q=3$, the $3$-states Potts mode, the twice-subleading magnetization and the $3$-cluster degenerate,
 giving $\theta^{(3)}_V|_{q=3}=\theta^{(3)}_Z|_{q=3}$.
 \item For $q=4$ the subleading $2$-cluster and the $3$-cluster degenerate,
 giving $\theta^{(3)}_T|_{q=4}=\theta^{(3)}_Z|_{q=4}$. 
\end{itemize}
Illustration of the degeneracies of the quadratic and cubic sectors are given in Figs.~\ref{fig:theta2s-d6}
and \ref{fig:theta3s-d6}, respectively.
Many more degeneracies can be expected by going beyond in the spectrum of irrelevant operators, but the above summarize all
the ones that we can verify with our computation and $q\geq 1$.
Degeneracies for the spanning tree and forest model at $q=0$ are more delicate to observe.
It has been argued that in this limit the identity operator (and consequently its subleading corrections, which coincide with the energy) degenerate with the magnetization upon careful normalization of the
factors of powers of $q$ in the limit. It is arguably difficult to observe the degeneracy of identity and
magnetization in the perturbative framework outside $d=2$,
since it would go across energy levels by literally following the discussion of \cite{Vasseur:2013baa}.
We notice however that $\theta^{(3)}_Z|_{q\to0}=0$ which agrees with $\theta^{(3)\prime}_V|_{q\to0}=0$,
instead of the expected $\theta^{(3)\prime}_V|_{q\to0}\neq0$.

In principle, each degeneracy leads through analytic continuation to a logarithmic CFT because the
Jordan cell has to be diagonalized.
In this case the coefficients $\alpha$ of the logarithmic contributions
arises from the further diagonalization and are universal (upon normalization with the traditional non-logarithmic contribution).
The coefficients can be determined as the limit of the finite difference
\begin{eqnarray}
 \alpha^{(M)}_X &=& \left.\frac{\Delta^{(M)}_{X}-\Delta^{(M)}_{\tilde{X}}}{q-q_c}\right|_{q\to q_c} 
 = \left.\frac{\theta^{{(M)}}_{\tilde{X}}-\theta^{{(M)}}_{X}}{q-q_c}\right|_{q\to q_c} 
 \label{alphadef}
\end{eqnarray} 
in which $X$ and $\tilde{X}$ are the spins of the two degenerate operators and $M$ is the order, as before.
Above we have also used the relation $\Delta^{{(M)}}_{X}=d-\theta^{{(M)}}_{X}$ to determine the operator's scaling dimension,
which strictly coincides with the CFT scaling dimension only for non-descendant operators.
We have for percolations, $q=1$, that $X=S$ and $\tilde{X}=T$, resulting in
\begin{equation}
\begin{split}
 \alpha^{(2)}_S &= -\frac{2 \epsilon }{7}
 +\frac{23 \epsilon ^2}{6174}
 +\left(-\frac{48 \zeta_3}{2401}+\frac{220751}{10890936}\right) \epsilon ^3 \,,
\\
 \alpha^{(3)}_S &= \frac{10 \epsilon }{21}
 -\frac{3403 \epsilon^2}{6174}
 +\left(\frac{1788 \zeta_3}{2401}+\frac{26582323}{32672808}\right) \epsilon ^3 \,.
\end{split}
\end{equation} 

Few other simple, but arguably less interesting,
logarithmic coefficients can be derived from our expressions in the case of $3$- and $4$-states Potts model.
For $q=3$, we have $X=V$ and $\tilde{X}=Z$, resulting in
\begin{equation}
\begin{split}
 \alpha^{(3)}_V &= \frac{90 \epsilon }{29}
 -\frac{9719 \epsilon ^2}{1682}
 +\left(\frac{120384 \zeta_3}{841}+\frac{109431967}{585336}\right) \epsilon ^3 \,.
\end{split}
\end{equation} 
For $q=4$, we have $X=T$ and $\tilde{X}=Z$, resulting in
\begin{equation}
\begin{split}
 \alpha^{(3)}_T &= -\frac{5 \epsilon }{6}
 -\frac{283 \epsilon ^2}{72}
 -\frac{193483 \epsilon ^3}{7776} \,.
\end{split}
\end{equation}

\section{In depth example: $q=4$} \label{sect:example}

It is interesting to examine more a special finite value of $q$, for which no analytic continuation is necessary,
to illustrate in practice several features of the spectrum of quadratic deformations
that we observed in the computations of the previous sections by comparing them with known literature's result.
For this purpose, we choose the case $q=4$, because it can be discussed in some detail,
although we have conducted similar extensive checks for all the values $5\leq q \leq 9$.
For the comparison,
we take appendix A of Ref.~\cite{Codello:2019isr}, in which the $4$-states Potts field theory
was discussed as a special case of a theory with general cubic interaction $\lambda_{ijk}\varphi_i\varphi_j\varphi_k$
without assuming $S_4$ symmetry a priori, but rather using only covariance of the potential under the maximal symmetry
group $O(3)$.
The obvious advantage of discussing the most general interaction is that deformations of the potential are not restricted by
the symmetry group of the potential itself, so we can observe how they arrange into $S_4$ multiplets.

The most general (symmetric) quadratic deformations $\varphi_i\varphi_j$ includes six distinct combinations,
$\varphi_1^2$, $\varphi_2^2$, $\varphi_3^2$, $\varphi_1 \varphi_2$, $\varphi_1 \varphi_3$ and $\varphi_2 \varphi_3$,
that have three distinct critical exponents, denoted $\theta^{(2)}$ in \cite{Codello:2019isr}, as relevant deformations
of the $4$-states Potts model at criticality. 
These correspond to the operators $S^{(2)}$, $V^{(2)}$ and $T^{(2)}$ of Sect.~\ref{sect:deformations},
which appear with the correct multiplicities of the $S_4$ irreducible representations,
as can be evinced by substituting $q=4$ in Eqs.~\eqref{eq:quad-deformations-d6}, \eqref{eq:quad-deformations-d6-2}
and \eqref{eq:quad-deformations-d6-3}.
Notice that the vector has multiplicity three because it has three independent components,
while the tensor has multiplicity two.

The situation becomes more complicate when discussing the symmetric cubic deformations $\varphi_i\varphi_j \varphi_k$
denoted $\theta^{(3)}$.
There are a total of ten distinct combinations that have four distinct critical exponents in Ref.~\cite{Codello:2019isr}.
Direct inspection reveals that one exponent correspond to the singlet $S^{(3)}$, and two are conjugate exponents corresponding to the mixing vectors $V^{(3)}$ and $V^{(3)\prime}$.
The remaining deformations arrange in a multiplet of three components that share the same critical exponent
$\theta^{(3)} \sim {\cal O}(\epsilon^2)$. In this case one can notice the degeneracy $\theta^{(3)}_T=\theta^{(3)}_Z$ for $q=4$, as one can check from Eqs.~\eqref{theta3T} and ~\eqref{theta3Z}, but keeping in mind that  in this case the $Z$ sector is not really physical by itself since it would be characterized by a negative dimensional space (see Eq.~\eqref{dim-decomp}), just opposite to the $T$ sector. 
This is of course the reason why the $T^{(3)}$ and $Z^{(3)}$ operators combine in the analytic continuation to $q\to 4$
resulting in a Jordan pair and a corresponding log-CFT.
Since in this paper we arranged deformations according to the irreducible representations of $S_{q}=S_{N+1}$,
while in \cite{Codello:2019isr} we arranged them according to the irreducible representations of $O(N)$,
we deduce that the additional vector does not transform according to any representation of $S_{q} \subset O(N)$,
but rather carries a representation label of $O(N)$.
This can be checked explicitly by looking at the components of the deformation.

The interesting aspect of the above discussion is that the deformation that does not carry a
label of the symmetry group $S_{q}$ of the
critical point has critical exponent with expansion starting at ${\cal O}(\epsilon^2)$,
in other words at the NLO of the $\epsilon$-expansion.
We conjecture here that this might be a general feature, namely that if a deformation of a critical point carries no irreducible
representation label of the symmetry of the critical point itself,
such deformation has a critical exponent with expansion that starts at least at ${\cal O}(\epsilon^2)$.
It would be interesting to test these features using a non-perturbative approach to the RG such as \cite{Zinati:2017hdy},
which allows to work directly in a physically interesting dimension, such as $d=2$ or $d=3$.

Finally, the analysis presented in \cite{Vasseur:2013baa} and verified in this paper suggests that
for $q=4$ the operators $T^{(3)}$ and $Z^{(3)}$ take pair of the same Jordan cell and for a logarithmic pair.
In fact, for $q=4$ they do not have enough components to form actual tensors
and their scaling dimensions do not show in the analysis of \cite{Codello:2019isr}.
In order to see them, it would be necessary to construct opportune observables as done for the case $q=1$
in Ref.~\cite{vjs,Gori:2017cyq}.

\bigskip

\section{RG estimates from $d=4-\epsilon$} \label{sect:estimates-d4}

In $d=4-\epsilon$ there are several universality classes of field theories with multiple scalar fields.
The discussion of these models is both old \cite{Brezin:1973jt,Zia:1974nv,Michel:1983in,PhysRevB.31.7171}
and new \cite{Osborn:2017ucf,Rychkov:2018vya},
and their general classification is still incomplete \cite{Rychkov:2018vya,Codello:2020lta},

The hypertetrahedral model \eqref{eq:quartic-potential} in $d=4-\epsilon$ dimensions
has received considerably less attention over the years,
both among its peers and compared to the sibling in $d=6-\epsilon$.
The study of the renormalization group flow of the couplings $(u,v)$ reveals three nontrivial fixed points.
Two are not of interest for us, because they do not have $S_q$ symmetry:
one is the well-known $O(N)$-symmetric fixed point of Wilson-Fischer,
and the other one that represents $N$ decoupled $\phi^4$ models, although in a parametrization that does not make it
straightforward to see and might require inspection of the spectrum. In both cases, the number of fieds is $N=q-1$.
This situation is very reminiscent of what is found when looking at the analog system
for the hypercubic model~\cite{Antipin:2019vdg,Zinati:2020xcn}.\footnote{In fact,
the two models even coincide for $q=4$,
because the symmetry group of the restricted Potts model coincides with the one of the cubic model,
$\mathbb{Z}_2 \times S_4 \simeq (\mathbb{Z}_2)^3 \rtimes S_3 =H_3$ \cite{Kousvos:2019hgc,Kousvos:2018rhl}.}
The last fixed point is the one with genuine $\mathbb{Z}_2\times S_q $ symmetry, which is thus named hypertetrahedral.

We computed all the same quantities for the hypertetrahedral model that we computed in the previous section,
starting from the RG in $d=4-\epsilon$ at next-to-next-to-leading order.
For illustrative purposes, we display it here at the leading order
\begin{eqnarray} \label{eq:beta-potential-d4}
\begin{split}
 \beta_v &= -d v + \frac{d-2}{2}  v_i \varphi_i +  v_i \gamma_{ij} \varphi_i +\frac{1}{2}  v_{ij}v_{ij}\,, \\
 \gamma_{ij} &= \frac{1}{12}v_{iklm}v_{jklm}\,,
\end{split}
\end{eqnarray}
for which summation over repeated indices is assumed.

\begin{figure}[htb]
  \includegraphics[width=0.49\textwidth]{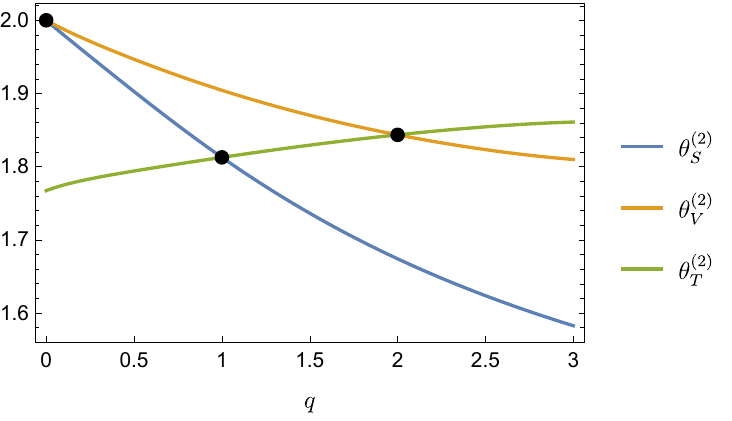}
  \caption{\label{fig:theta2s-d4} Plot of the critical exponents $\theta^{(2)}_X$ for $X=S,V,T$ extrapolated to $d=3$
  ($\epsilon=1$) of the hypertetrahedral model as a function of $q$. The marked points highlight the degeneracies
  that lead to logarithmic behavior.}
\end{figure}

We want to report briefly some of such quantities, because, to the best of our knowledge,
most of them have not been given elsewhere.
We still denote with $\theta^{(n)}_X$ the critical exponent associated with the relevant deformation containing $n$ fields and irreducible multiplet $X=S,V,T,Z$, which should not be confused with those given in the
case of the Landau-Potts field theory. We find for the $n=2$ sector
\begin{widetext}
\begin{eqnarray}
 \theta^{(2)}_S &=& 
  2
  -\frac{2 q \epsilon }{3 (q+2)}
  +\frac{q \left(19 q^3-258 q^2+1332 q+648\right) \epsilon ^2}{162 (q-6) (q+2)^3}
  +\epsilon ^3
   \Bigl\{
  -\frac{4 q \left(q^4-3 q^3-20 q^2+228 q+144\right) \zeta_3}{27 (q-6) (q+2)^4}
 \\&&
  +\frac{q \left(937 q^7+28770 q^6-463356 q^5+2772360 q^4-8850384 q^3+11018592
   q^2+9471168 q+6998400\right)}{17496 (q-6)^3 (q+2)^5}
\Bigr\}\,,
 \nonumber 
\end{eqnarray}
\begin{eqnarray}
 \theta^{(2)}_V &=& 
  2
  -\frac{q \epsilon }{3 (q+2)}
  + \frac{q \left(-19 q^3-132 q^2+828 q+432\right) \epsilon ^2}{162 (q-6) (q+2)^3}
  +\epsilon ^3
   \Bigl\{
  \frac{4 q \left(q^4+6 q^3-5 q^2-150 q-72\right) \zeta_3}{27 (q-6) (q+2)^4}
  \\&&
  +\frac{q \left(-937 q^7+6966 q^6-77220 q^5+974808 q^4-5519664 q^3+7488288 q^2+8071488
   q+5318784\right)}{17496 (q-6)^3 (q+2)^5}
  \Bigr\}\,,
 \nonumber 
\end{eqnarray}
\begin{eqnarray}
 \theta^{(2)}_T &=& 
 2+
 -\frac{2 \epsilon }{3 (q+2)}
 +\frac{\left(-3 q^4+124 q^3-252 q^2+1008 q+864\right) \epsilon ^2}{162 (q-6) (q+2)^3}
 -\epsilon^3 \Big\{
 \frac{8 \left(2 q^4+6 q^3-7 q^2+102
   q+72\right) \zeta_3}{27 (q-6) (q+2)^4}
 \\&&
 +\frac{-327 q^8+4354 q^7+93972 q^6-1099512 q^5+4173552 q^4-6192288 q^3+5800896
   q^2+11477376 q+6718464}{17496 (q-6)^3 (q+2)^5}
 \Bigr\}\,.
 \nonumber 
\end{eqnarray}
%
As before we include the anomalous dimension for completeness
\begin{eqnarray}
 \eta &=&
  \frac{q (q+6) \epsilon ^2}{54 (q+2)^2} 
  +\frac{q \left(109 q^3+642 q^2+3852 q+2808\right) \epsilon ^3}{5832 (q+2)^4}\,.
\end{eqnarray}
Some degeneracies of the quadratic sector are displayed in Fig.~\ref{fig:theta2s-d4}

Notice that the plot highlights a further degeneracy, involving the scalar and vector quadratic deformations,
that is both interesting and trivial. It is interesting because the scalar and vector deformations
are to be degenerate, which is arguably not easy to see in the $\epsilon$-expansion. However, the result is trivial,
in the sense that it is mean-field,
because $2=\theta_S^{(2)}=\theta_V^{(2)}$ for $q\to 0$. 
Neverthelss, the universal coefficient of the logarithm associated to this degeneracy is non trivial
and given by ($X=S$ and $\tilde X=V$ in Eq.~\eqref{alphadef})
\be
\alpha_S^{(2)}=
\frac{\epsilon}{6} +\frac{\epsilon ^2}{36}-\left(\frac{\zeta_3}{9}-\frac{1}{72}\right) \epsilon ^3\,.
\ee
Even though this degeneracy was expected on the basis of the continuation to $q\to 0$ of the representation
theory of $S_q$
(see the relevant discussion appearing in \cite{Vasseur:2013baa}),
we did not observe it in Sect.~\ref{sect:estimates-d6},
so it is amusing to see it here.
In fact, this is the only logarithmic coefficient that we display for the reduced Landau-Potts model,
especially because it might be relevant
for applications to the random cluster model \cite{Deng:2006ur}.
We stress, however, that analog universal coefficients for the other two degeneracies (the same that are shown in Fig.~\ref{fig:theta2s-d4})
can be obtained in a similar way.

The expressions are easy to specialize to obtain the fractal dimensions as before.
We could still refer to the limits $q\to0$ and $q\to1$ as spanning forest and percolation, respectively,
however, to our knowledge, the potential relations of these limits with the mentioned models
has not been explored in the case of the hypertetrahedral model.
In the limit $q\to 0$ we have
\begin{equation}
\begin{split}
 d_{r,q=0} = 2-\frac{\epsilon }{3} -\frac{\epsilon^2}{9}
  +\left(\frac{2 \zeta_3}{9}-\frac{1}{18}\right) \epsilon ^3
 \,, \qquad
 d_{f,q=0} = 2
 \,,
\end{split}
\end{equation}
while in the limit $q\to 1$
\begin{equation}
\begin{split}
 d_{r,q=1} &= 2 -\frac{2 \epsilon }{9} -\frac{1741 \epsilon ^2}{21870}
 +\left(\frac{280 \zeta_3}{2187}-\frac{20976487}{531441000}\right) \epsilon ^3 \,,
 \\
 d_{f,q=1} &= 2 -\frac{\epsilon }{9}-\frac{1109 \epsilon ^2}{21870}
 +\left(\frac{176 \zeta_3}{2187}-\frac{16262513}{531441000}\right) \epsilon ^3 \,.
\end{split}
\end{equation}

The crossover exponents are defined as before as the ratio of critical exponents. For the case $n=2$ we find
\begin{eqnarray}
 \overline{\Phi} &=& 1 +\frac{q \epsilon }{6 (q+2)}
  -\frac{q \left(10 q^3-27 q^2+360 q+108\right) \epsilon ^2}{162 (q-6) (q+2)^3}
  +\epsilon ^3
   \Bigl\{\frac{2 q \left(2 q^4+3 q^3-25 q^2+78 q+72\right) \zeta_3}{27 (q-6)
   (q+2)^4}
   \\&&
   -\frac{q \left(734 q^7+1149 q^6-64404 q^5+328860 q^4-850176 q^3+1839024 q^2+699840
   q+419904\right)}{8748 (q-6)^3 (q+2)^5}\Bigr\}
 \nonumber
 \\
 \Phi &=&
  1+\frac{(q-1) \epsilon }{3 (q+2)}
  \frac{\left(7 q^3+143 q^2-78 q-72\right) \epsilon ^2}{162 (q+2)^3}
 +\epsilon ^3 \Bigl\{
  \frac{2\left(q^4-q^3-38 q^2+14 q+24\right) \zeta_3}{27 (q+2)^4}
 \\&&
  -\frac{361 q^7+3509 q^6-43470
   q^5+333792 q^4-1070496 q^3+99792 q^2+396576 q+279936}{8748 (q-6)^2
   (q+2)^5}\Bigr\}
 \nonumber
\end{eqnarray} 
\end{widetext}
We derived similar unwieldy expressions for all the cubic operators, which we refrain from giving here because of space.
All complete beta functions and critical exponents are provided through an ancillary file.

We have also explicitly checked that all expected patterns of degeneracy in the spectrum are still present.
The reason is that, following the logic of Refs.~\cite{Vasseur:2013baa,vjs}, these degeneracies are related to the symmetry of the model,
and are thus shared between the Landau-Potts and hypertetrahedral field theories.
This suggest the possibility of constructing several more log-CFTs as limits in $d=4-\epsilon$ dimensions.
In this respect, the computation of the universal coefficients of the logarithms follows exacly
the same steps as before.

We conclude by giving some numerical estimates for few interesting limits of $q$.
We estimate all the critical quantities in $d=3$ by resumming the $\epsilon$-expansion with the Pad\'e-Borel method
and taking $\epsilon=1$.
We use Pad\'e approximants $(2,1)$ and $(1,2)$ for the Borel transform whenever possible
(if the series does not allow either approximant, we replace it with $(1,1)$).
Errors are estimated by inspecting the difference of the available resummations in order to give
a rough idea, and therefore error bars are not at all meant to
be exact bounds or to suggest a convergence scheme.
Estimates for the limits $q\to 0$ and $q\to 1$ are given in Tab.~\ref{tab:table1} and Tab.~\ref{tab:table2}, respectively.
\begin{table}[h!]
  \begin{center}
    \begin{tabular}{|c|c|c|} 
	\hline 
	$d_r$ 		& $\Phi$ 		& $\alpha_S^{(2)}$ 	\\
	\hline
	$1.64(4)$ 	&  $0.82(2)$ 	& $ 0.16(3)$		\\
	\hline
    \end{tabular}
    \caption{Pad\'e-Borel estimates for the spanning forest limit $q=0$ in $d=3$.
    All other exponents can be obtained exactly using $q=0$ in the formulas of this section.}
    \label{tab:table1}
  \end{center}
\end{table}
\begin{table}[h!]
  \begin{center}
    \begin{tabular}{|c|c|c|c|} 
	\hline 
	$\nu^{-1}$ 	& $d_f$ 		& $d_r$ 		& $\alpha_S^{(2)}$ 	\\
	\hline
	$1.75(2) $	& $1.86(1)$	& $1.75(2)$ 	&  $0.23(6)$		\\
	\hline
    \end{tabular}
    \caption{Pad\'e-Borel estimates for the percolation limit $q=1$ in $d=3$.
    The fractal dimensions $d_f$ and $d_r$ are the critical exponents $\theta_V^{(2)}$ and $\theta_T^{(2)}$.
    The corresponding operator scaling dimensions
    can be easily obtained using the relation $\Delta=d-\theta$, which holds for primary CFT operators.}
    \label{tab:table2}
  \end{center}
\end{table}

We also give estimates for the case $q=4$ in Tab.~\ref{tab:table3},
because the limit coincides with the cubic model. The operators $V$ and $T$ of
this paper are identified with $X$ and $Y$ of \cite{Kousvos:2019hgc,Kousvos:2018rhl}
(see also \cite{Antipin:2019vdg,Zinati:2020xcn} for the discussion of the underlying irreps).
The combined numerical effort gives
the scaling dimensions $\Delta_\varphi \approx 0.520$, $\Delta_S \approx 1.521$, $\Delta_X\approx 1.221$,
and $\Delta_Y \approx 1.181$. These numbers seem to indicate that the cubic model lies in the allowed region of the bootstrap bounds of \cite{Stergiou:2018gjj,Rong:2017cow}.
\begin{table}[h!]
  \begin{center}
    \begin{tabular}{|c|c|c|c|} 
	\hline 
	$\eta$ 		& $\nu^{-1}$	& $d_f$ 		& $d_r$ 			\\
	\hline
	$0.039^* $	& $1.479(56)$	& $1.779(16)$ 	&  $1.819(17)$		\\
	\hline
    \end{tabular}
    \caption{Pad\'e-Borel estimates for the cubic model as the limit $q=4$ in $d=3$.
    The anomalous dimension comes from the extrapolation to $\epsilon=1$ of the $\epsilon$-expansion
    ($\eta$ is marked with an asterisk because it is obtained by extrapolation without resummation).
    The fractal dimensions $d_f$ and $d_r$ are the critical exponents
    of the tensor operators $X$ and $Y$ of the cubic model, so they can be used to determine the estimates in the main text
    using the relation $\Delta=d-\theta$.}
    \label{tab:table3}
  \end{center}
\end{table}

\section{Conclusions}

The analysis of the critical point of the $q$-states Potts model can be done using the standard 
Landau-Ginzburg approach, which we have applied to three loop order to obtain next-to-next-to-leading
order exponents in the $\epsilon$-expansion. While $\epsilon$-expansion converges rather poorly
for interesing values of lower dimensions of interest such as two or three, it is still a very convenient tool
when it comes to unveiling universal properties. One such property is the dependence on the parameter $q$ of critical
exponents. We have computed the exponents associated to non-singlet deformations of the Landau-Ginzburg potential
of second and third order in the field.

In short, our work adds the next-to-next-to-leading order in the computations
of non-singlet deformations of the Landau-Potts field theory
with $\phi^3$-type interaction, discusses the same order for the 
restricted Potts field theory with $\phi^4$-type interaction (also known as the restricted Potts model),
and combines all the results to provide estimates of crossover exponents,
fractal dimensions of propagator lines and new universal quantities related to logarithmic CFTs.

Non-singlet deformations are interesting because they are related to statistically and geometrically meaningful observables, thus
allowing to estimate the corresponding critical exponents. In particular, we have estimated critical crossover exponents
for the breaking of the $S_q$ symmetry when the temperature is at its critical value. In a similar fashion, we have
computed the expansion of the fractal dimensions associated to Potts clusters. Finally, we used the full power of
the $q$-dependence of our estimates to discuss the emergence of logarithmically scaling operators
in the underlying CFT. The discussions on logarithms is particularly interesting because it shows that
there are new universal quantities, the coefficients of such logarithms, that can be computed by means of RG methods
and that require a complete handle on the $q$-dependence of the spectrum.

On this note, having accepted the limitations of the $\epsilon$-expansion in this particular context,
one important point that our paper shows is that several interesting quantities,
CFT's logarithms among all, are accessible within RG techniques. This implies that they could be computed by RG methods
other than dimensional regularization to improve the value of the corresponding estimates.
This is an interesting possibility in the light that recent works have shown that the logarithmic effects are
visible when studying the scaling of relatively simple observables~\cite{vjs,Gori:2017cyq}.

\smallskip

\paragraph*{Acknowledgments}
The work presented in this paper was inspired
by a relatively recent paper by Kompaniets and Wiese,
which studied similar quantities for the $O(N)$ model \cite{Kompaniets:2019zes}.
For performing the computations,
we relied heavily on the Mathematica packages \cite{xact-package,xperm-package} and \cite{Nutma:2013zea}.
OZ is grateful to R.~Ben Al\`i Zinati for several discussions on the topics covered by this work.
We are grateful to the anonymous referee for suggestions that improved the quality of the draft.

\bibliography{bibliography}

\end{document}